\documentclass[twocolumn,prb]{revtex4}
\usepackage{amsfonts}
\usepackage[T1]{fontenc}
\usepackage{amsmath,amsbsy,amssymb,graphicx}
\usepackage{times}

\begin{document}

\title{Minimal model for higher-order topological insulators and phosphorene}
\author{Motohiko Ezawa}
\affiliation{Department of Applied Physics, University of Tokyo, Hongo 7-3-1, 113-8656,
Japan}

\begin{abstract}
A higher order topological insulator is an extended notion of the
conventional topological insulator, which belongs to a special class of
topological insulators where the conventional bulk-boundary correspondence
is not applicable. The bulk topological index may be described by the
Wannier center located at a high symmetry point of the crystal. In this
paper we propose minimal models for the second-order topological insulator
in two dimensions and the third-order topological insulator in three
dimensions. They are anisotropic two-band models with two different hopping
parameters. The two-dimensional model is known to capture the essential
physics of phosphorene near the Fermi level. It has so far been recognized
as a trivial insulator due to the absence of topological edge states in
nanoribbons. However, we demonstrate the emergence of topological boundary
states in zero dimension, i.e., in nanodisks. In particular, the diamond
structure exhibits such topological states at two corners, each of which
carries a 1/2 fractional charge. We predict that these corner states will be
observed in the diamond structure of phosphorene.
\end{abstract}

\maketitle

A topological insulator (TI) is characterized by the bulk topological index
together with the emergence of topological boundary states\cite{Hasan,Qi}.
The gap must close along the boundary since the topological index cannot
change its value continuously across the boundary, which is known as the
bulk-boundary correspondence. The typical bulk topological index is the
Chern number, which is the genuine one. Another typical one is the Z$_{2}$
index, which is protected by the time-reversal symmetry. However, recently,
the concept of the TI has been generalized to include the higher-order TI
(HOTI)\cite{Fan,Science,APS,Peng,Lang,Song,Bena,Schin,Liu,EzawaKagome}. Let
us consider a $j$-dimensional ($j$D) bulk system. For instance, a
second-order TI is an insulator which has $(j-2)$D topological boundary
states but no $(j-1)$D topological boundary states. Similarly, a third-order
TI is an insulator which has $(j-3)$D topological boundary states but no $(j-1)$D 
and $(j-2)$D topological boundary states. Namely, the boundary of
the third-order TI is the second-order TI. In spite of these properties, the
HOTI is also characterized by the bulk topological index\cite{Science,Bena,Song,EzawaKagome}. 
It belongs to a special class of TIs to
which the conventional bulk-boundary correspondence is not applicable.
Accordingly, there must be HOTIs which have so far been regarded as trivial
insulators.

Recent studies have revealed a crucial role that the Wannier center (WC)
plays in a certain type of HOTIs\cite{Science,Bena,Song,EzawaKagome}. The WC
is the expectation value of the position in the unit cell of a crystal and
given by the integral of the Berry connection over the Brillouin zone. The
WC is fixed at the high symmetry point. Since it is proved that its value is
quantized in insulators, it can be used as a new type of the topological
number\cite{Science,Bena,Song,EzawaKagome}. We consider the atomic insulator
limit, where there are no interactions among any lattice sites. The WC is on
the lattice site, and the system is a trivial insulator. In general, it is
trivial when the WC is on the lattice site, and topological otherwise\cite{Po,Po2,Po3,Brad}.

Here we propose the minimal models of Wannier-type HOTIs in two and three
dimensions. We first analyze a 2D crystal (anisotropic honeycomb lattice)
possessing the chiral symmetry as the minimal model. In this model we show
that the WC is identical to the winding number. We calculate the energy
spectra for the bulk, the nanoribbon\cite{Nanoribbon} and the nanodisk\cite{Nanodisk} 
to search for zero-energy states. Then, we point out that
phosphorene is a second-order HOTI, although it has so far been regarded as
a trivial insulator due to the absence of topological boundary states. This
may well be the first example of HOTIs whose samples are already available
in laboratories. It is also possible to construct a 3D crystal (anisotropic
diamond lattice) so that its boundary is the above 2D crystal (anisotropic
honeycomb lattice) when we cut it along a plane. Then, it is a third-order
TI by definition. These anisotropic lattice models are higher-dimensional
extensions of the SSH model.

\medskip\noindent\textbf{\large Minimal two-band models}

HOTIs were originally proposed on square lattices with the use of the
four-band model\cite{Fan,Science,APS,Peng,Lang,Song,Bena,Schin,Liu}, and
subsequently studied on the breathing Kagome and pyrochlore lattices with
the use of the three-band and four-band models\cite{EzawaKagome},
respectively. The minimal model to describe insulators is the two-band
model. We propose a two-band model to reveal the essence of HOTIs.

We consider lattice models which have two atoms in the unit cell. By
requiring the chiral symmetry, the Hamiltonian is described by the two band
theory
\begin{equation}
H_{0}=\left( 
\begin{array}{cc}
0 & F \\ 
F^{\ast } & 0
\end{array}
\right) ,  \label{H0}
\end{equation}
where the chiral symmetry operator is $C=\sigma _{z}$ with $C^{-1}H_{0}C=-H_{0}$. 
The diagonal terms are prohibited by the chiral
symmetry. The energy spectrum reads $E=\pm \left\vert F\right\vert $, which
is symmetric ($E\leftrightarrow -E$) due to the chiral symmetry. Typical
examples are the SSH model in one dimension, the anisotropic honeycomb
lattice in two dimensions and the anisotropic diamond lattice in three
dimensions. For the SSH model\cite{SSH}, we take
\begin{equation}
F=t_{a}+t_{b}e^{ik}.
\end{equation}
For the anisotropic honeycomb lattice, we take\cite{Wunsch,Goerbig,Pereira,Monta,EzawaPhos}
\begin{equation}
F=2t_{a}e^{-ik_{y}/2}\cos \frac{\sqrt{3}k_{x}}{2}+t_{b}e^{ik_{y}}.
\end{equation}
For the anisotropic diamond lattice, we take\cite{Takahashi}
\begin{equation}
F=t_{a}\left( e^{i\mathbf{k}\cdot \mathbf{X}_{2}}+e^{i\mathbf{k}\cdot 
\mathbf{X}_{3}}+e^{i\mathbf{k}\cdot \mathbf{X}_{4}}\right) +t_{b}e^{i\mathbf{k}\cdot \mathbf{X}_{1}},
\end{equation}
with the four lattice vectors pointing the tetrahedron directions 
$\mathbf{X}_{1}=\left( 1,1,1\right) ,$ $\mathbf{X}_{2}=\left( 1,-1,-1\right) ,$ 
$\mathbf{X}_{3}=\left( -1,1,-1\right) $ and $\mathbf{X}_{4}=\left(-1,-1,1\right) $. 
We take $t_{b}>0$ without loss of generality.

\begin{figure}[t]
\centerline{\includegraphics[width=0.49\textwidth]{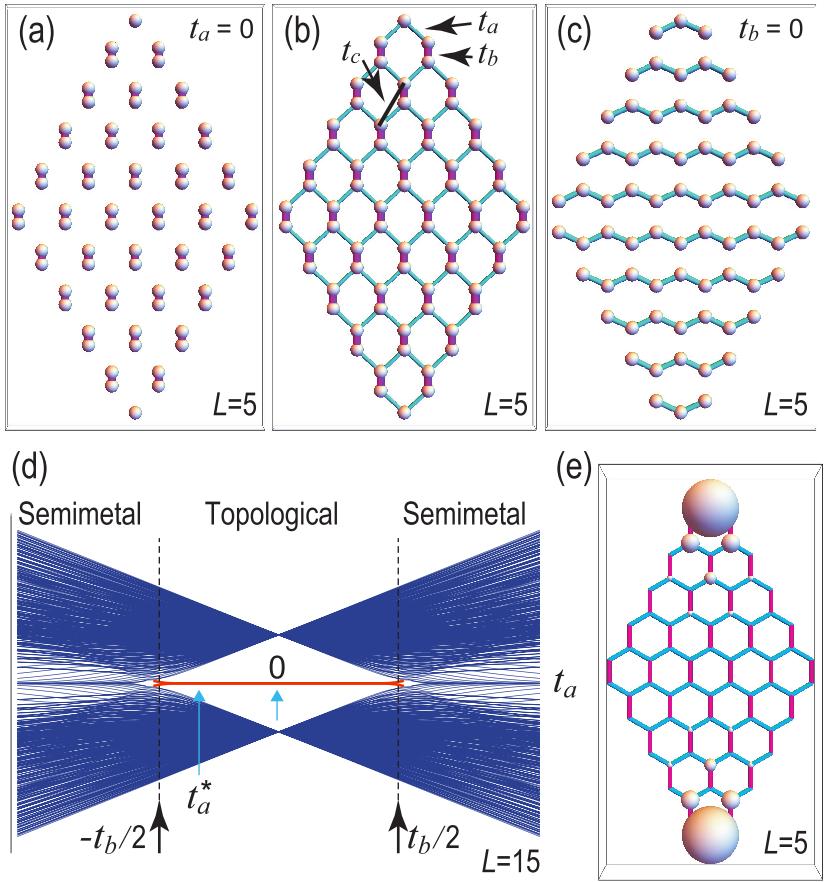}}
\caption{Illustration of diamonds with (a) $t_{a}=0$, (b) $\left\vert
t_{a}\right\vert <t_{b}$ and (c) $t_{b}=0$. The size $L$ is defined by the
number of Benzene rings on one side in (b). (a) There are two isolated atoms
at the top and bottom corners of the diamond for $t_{a}=0$. (b) All four
edges form skew-zigzag nanoribbons well described by the SSH model. The two
corners at the top and the bottom are topological zero-energy states
inherent to the SSH model. (d) Energy spectrum of the diamond made of the
anisotropic honeycomb lattice. The horizontal axis is $t_{a}$ with a fixed
value of $t_{b}=3.665$eV, where $t_{a}^{\ast }=-1.220$eV. There emerge
zero-energy states (marked in red) for $|t_{a}|<t_{b}/2$. They represent
topological boundary states. (e) The square root of the local density of
states $\protect\sqrt{\protect\rho _{i}}$ for the diamond. The amplitude is
represented by the radius of the spheres. The local density of states
becomes arbitrarily small except for the two corners for $L\gg 1$,
indicating that the localized states emerge only at the two corners of the
diamond. }
\label{FigDiamond}
\end{figure}

\begin{figure}[t]
\centerline{\includegraphics[width=0.48\textwidth]{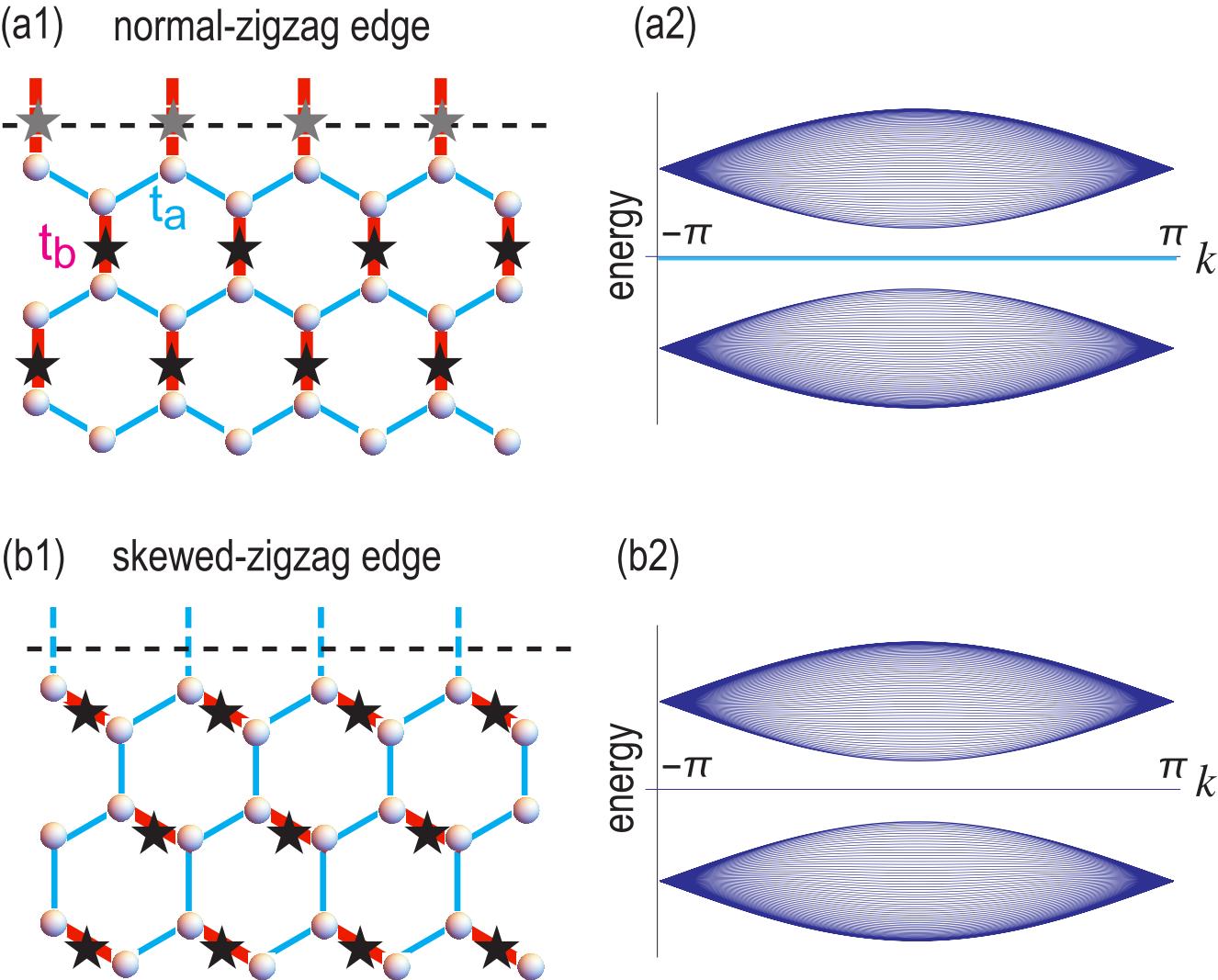}}
\caption{Illustration and band structure of (a1,a2) a normal-zigzag
nanoribbon and (b1,b2) a skew-zigzag nanoribbon with $t_{a}=-1.220$eV and $t_{b}=3.665$eV. 
Star symbols represent WCs in (a1,b1). The horizontal axis
is $k$ in (a2,b2). (a2) When the edge cuts through WCs as in (a1), a perfect
flat band emerges at zero energy as represented by a cyan line. This perfect
flat band is detached entirely from the bulk band, and it does not represent
a topological boundary state. (b2) When the edge cuts through no WCs as in
(b1), no boundary states emerge at zero energy. }
\label{Ribbon}
\end{figure}

In the case of the SSH model, the WC is given by the polarization $p_{x}$
along the $x$ axis, which is the bulk topological index protected by the
mirror symmetry along the $x$ direction. We generalize it to higher
dimensions\cite{Science,Bena,Song,EzawaKagome}. The WC is given by the set
of the $j$ polarization $p_{\alpha }$ in the $j$ dimensions, which is the
average of the position in the unit cell, 
\begin{equation}
p_{\alpha }=-\frac{1}{V}\int_{\text{BZ}}d^{j}kA_{\alpha },  \label{PolarP}
\end{equation}
where $A_{\alpha }=-i\left\langle \psi \right\vert \partial _{k_{\alpha
}}\left\vert \psi \right\rangle $ is the Berry connection, $V$ is the volume
of the Brillouin zone, and the integration is carried out over the Brillouin
zone. Due to the gauge invariance of the polarization\cite{Science}, 
$p_{\alpha }$ is defined mod $1$. Furthermore, in the presence of the mirror
symmetry, $M_{\alpha }^{-1}H_{0}\left( k_{\alpha }\right) M_{\alpha
}=H_{0}\left( -k_{\alpha }\right) $, the polarization is odd\cite{Science}, 
$p_{\alpha }\rightarrow -p_{\alpha }$, with respect to it. Combining these
two properties, we find\cite{Science,Bena,Song,EzawaKagome} that the
polarization $p_{\alpha }$ is quantized to be $0$ or $1/2$.

The topological nature of the polarization is made manifest for the chiral
symmetric two-band theory. Using the Hamiltonian (\ref{H0}) explicitly, we
may solve for the eigen function of the ground state as $\psi
=\left(-e^{-i\Theta },1\right) ^{t}/\sqrt{2}$ with $\Theta =i\log (F/\left\vert F\right\vert )$. The connection
reads $A_{\alpha}=-\frac{1}{2}\partial _{k_{\alpha}}\Theta$,
and hence the formula (\ref{PolarP}) represents the winding number.

We use the WC as the bulk topological number. We consider the atomic
insulator limit\cite{Brad,Po,Po2,Po3} by taking $t_{a}=t_{b}=0$, where there
are no interactions among any lattice sites, and we find $p_{\alpha }=0$
trivially. Hence, when $p_{\alpha }=0$ for all $\alpha $, it is natural to
define that the system is trivial, and otherwise topological\cite{Brad,Po,Po2,Po3}. 
As we shall show, the WC is $1/2$ for the SSH model, 
$\left( 0,1/2\right) $ for the anisotropic honeycomb lattice and 
$\left(1/2,1/2,1/2\right) $ for the anisotropic diamond lattice. They exist at the
center of the dimerized bonds. The WC cannot change its value as long as the
bulk band gap does not close as a function of $t_{\alpha }/t_{b}$.

\medskip \noindent \textbf{\large Phosphorene}

The anisotropic honeycomb lattice is naturally realized in phosphorene\cite{EzawaPhos}, 
which is a monolayer material of black phosphorus. In
phosphorene, there is an additional term $f$ to the anisotropic honeycomb
lattice model (\ref{H0}),
\begin{equation}
H=H_{0}+fI,  \label{Hphos}
\end{equation}
where $I$ is the $2\times 2$ identity matrix and
\begin{equation}
f=4t_{c}\cos \frac{\sqrt{3}}{2}k_{x}\cos \frac{1}{2}k_{y}.
\end{equation}
Three hopping parameters $t_{a}$, $t_{b}$ and $t_{c}$ are shown in Fig.\ref{FigDiamond}, 
which are $t_{a}=-1.220$eV, $t_{b}=3.665$eV, $t_{c}=-0.105$eV
according to Ref.\cite{Kats}. We have checked numerically that the neglected
terms do not affect the topological argument at all.

The diagonal term $fI$ breaks the chiral symmetry and modifies the perfect
flat band [Fig.\ref{Ribbon}(a2)] into the quasi-flat band characteristic to
phosphorene\cite{EzawaPhos}. This anisotropic two-band model may well
capture the essential physics of phosphorene near the Fermi level, where 
$t_{a}/t_{b}\approx -1/3$. Since the diagonal term $fI$ only shifts the
energy levels by $f$ without affecting the topological structure, it is
enough to analyze the Hamiltonian $H_{0}$ to reveal the topological aspect
of phosphorene.

\begin{figure}[t]
\centerline{\includegraphics[width=0.49\textwidth]{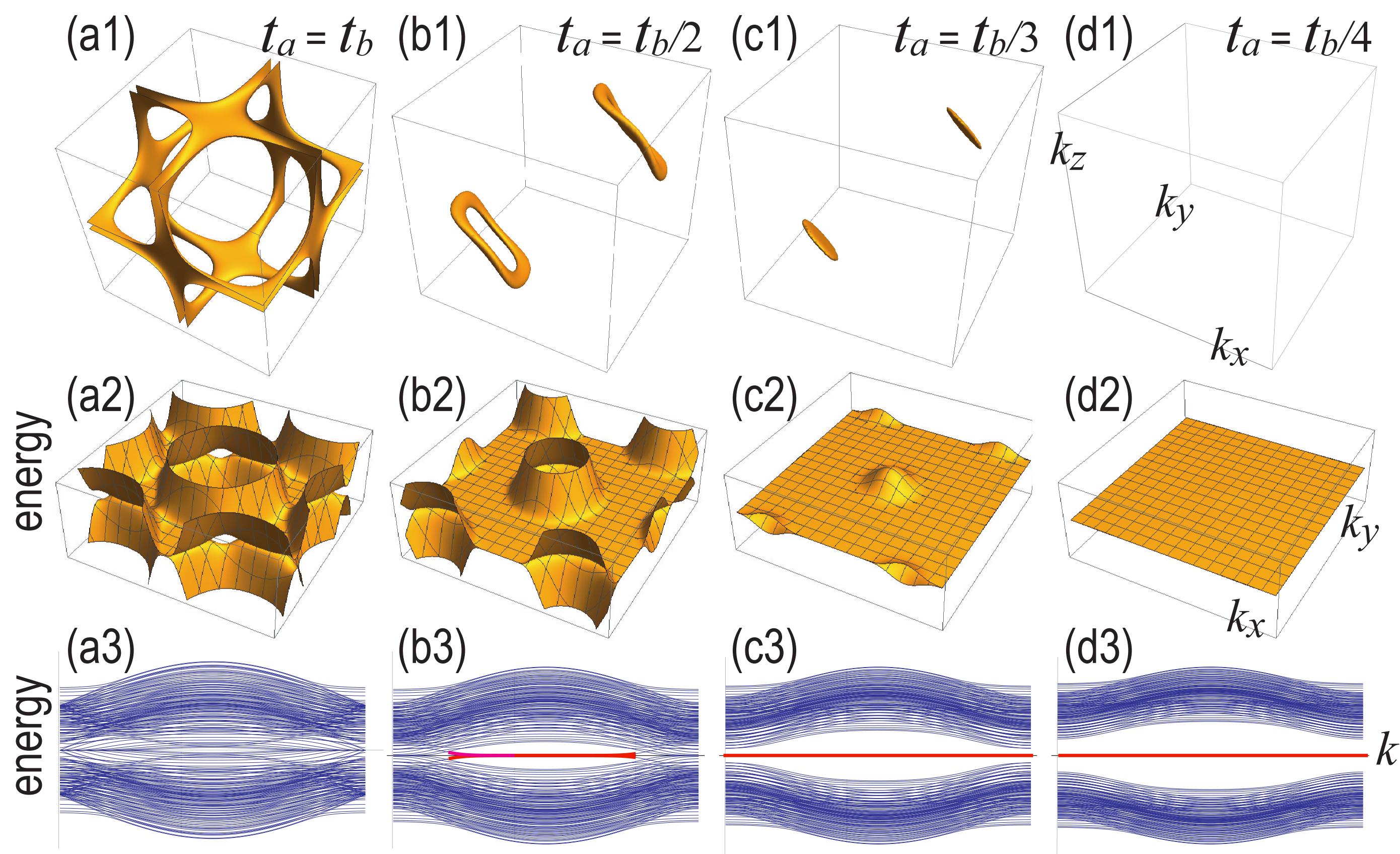}}
\caption{The Fermi surface of the bulk anisotropic diamond lattice with (a1) 
$t_a=t_b$, (b1) $t_a=t_b/2$, (c1) $t_a=t_b/3$ and (d1) $t_a=t_b/4$.
(a2)-(d2) The corresponding band structure of a thin film. (a3)-(d3) The
corresponding band structure of a diamond prism. The size of the diamond is $L=3$.}
\label{FigAniBand}
\end{figure}

We now show that the two-band model $H_{0}$ describes a second-order TI in
the range of parameters $|t_{a}|<t_{b}/2$ by making four-step arguments. We
calculate the energy spectra for the bulk, the nanoribbon and the nanodisk
in the first three steps. Finally we make the topological arguments.

(i) First, we examine the bulk band spectrum. The dispersion relation reads 
\begin{equation}
E=\pm \sqrt{t_{b}^{2}+4\left( t_{a}^{2}+t_{a}t_{b}\cos \frac{\sqrt{3}}{2}k_{x}\right) \cos \frac{k_{y}}{2}},
\end{equation}
Two Dirac cones exist at the $K$ and $K^{\prime }$ points $\left( 0,\pm k_{\text{D}}\right) $ with
\begin{equation}
k_{\text{D}}=2\arctan \left( \sqrt{4t_{a}^{2}-t_{b}^{2}}/t_{b}\right)
\end{equation}
for $\left\vert t_{a}\right\vert >t_{b}/2$ and the system is semimetallic.
The two Dirac cones merge at the $\Gamma $ point for $\left\vert
t_{a}\right\vert= t_{b} /2$, and the system becomes an insulator for $\left\vert t_{a}\right\vert <t_{b}/2$.

\begin{figure}[t]
\centerline{\includegraphics[width=0.49\textwidth]{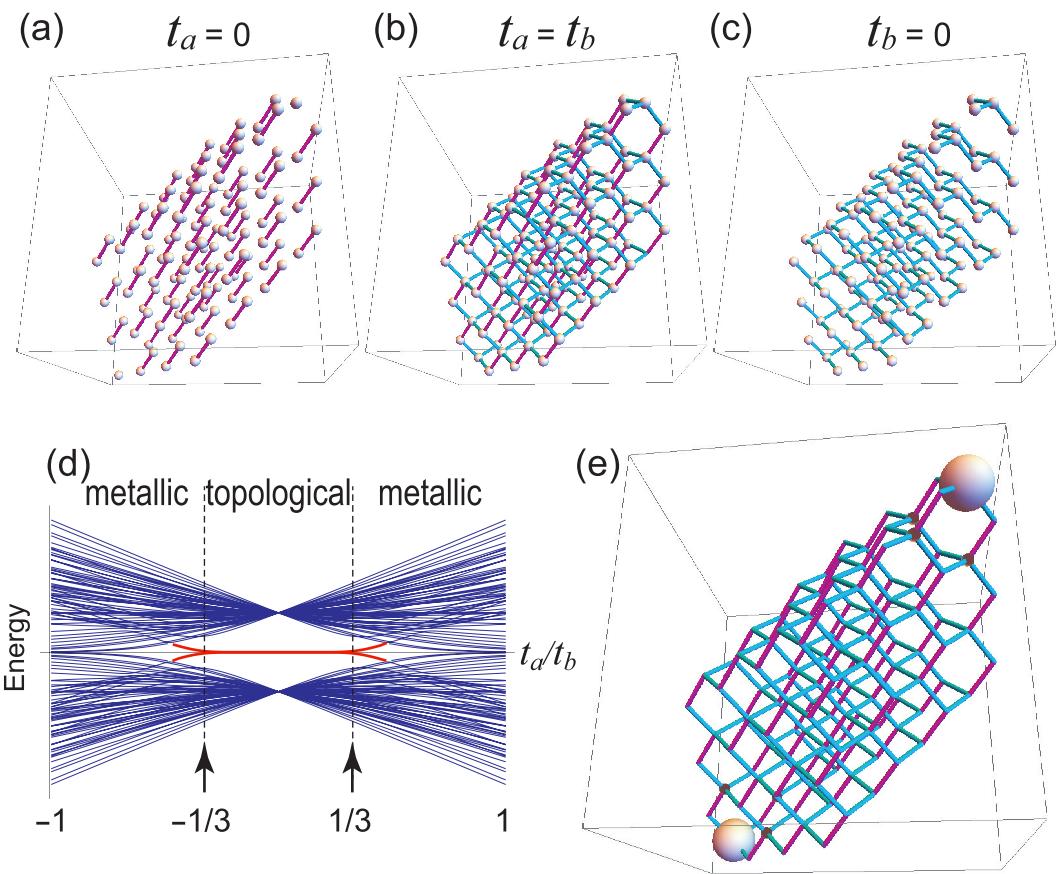}}
\caption{Illustration of a rhombohedron made of the diamond lattice with (a) 
$t_a=0$, (b) $t_a=t_b$ and (c) $t_b=0$. (d) Energy spectrum of the
rhombohedron made of the anisotropic diamond lattice with $L=3$. The
horizontal axis is $t_{a}/t_{b}$. There emerge zero-energy states (marked in
red) for $|t_{a}/t_{b}|<1/3$. They are topological boundary states. (e) The
square root of the local density of states $\protect\sqrt{\protect\rho _{i}}$
for the rhombohedron with $L=3$ and $t_{a}/t_{b}=1/4$. The amplitude is
represented by the radius of the spheres. The local density of states
becomes arbitrarily small except for the two corners for $L\gg 1$. The
localized states emerge only at the two corners of the rhombohedron. }
\label{FigAniDia}
\end{figure}

ii) Second, we study the energy spectrum of the 1D boundary. Let us cut a
crystal along a line to create a normal-zigzag edge, where the edge passes
through the WCs, as in Fig.\ref{Ribbon}(a1). Actually we analyze the band
structure of a normal-zigzag nanoribbon\cite{EzawaPhos}. There exists a
partial flat band connecting the $K$ and $K^{\prime }$ points for 
$\left\vert t_{a}\right\vert >t_{b}/2$. It becomes a perfect-flat zero-energy
edge state for $\left\vert t_{a}\right\vert <t_{b}/2$: See Fig.\ref{Ribbon}(a2). 
However, they are not topological edge states implied by the
conventional bulk-edge correspondence, since they are entirely detached from
the bulk band. The emergence of the zero-energy edge states is understood as
follows. The WC exists at the middle of the bond connecting the $A$ and $B$
sites in the topological phase: See Fig.\ref{Ribbon}(a). The charge
distributes in the vicinity of the WC. A half charge emerges at the boundary
since the charge is separated into two pieces by the boundary on the WCs,
yielding zero-energy boundary states protected by the chiral symmetry.

We may also study the energy spectrum of a skew-zigzag nanoribbon\cite{Gru}:
See Fig.\ref{Ribbon}(b1). A crucial difference is the absence of a
zero-energy edge state: See Fig.\ref{Ribbon}(b2). One edge of a nanoribbon
realizes a SSH chain. The absence of the zero-energy edge states is
understood as follows. The edge passes through no WCs when we cut a crystal
along a line to create a skew-zigzag edge, as in Fig.\ref{Ribbon}(b1). No
charge emerges at the edge since there are no charge at the boundary away
from the WCs.

iii) Third, we study the energy spectrum of the 0D boundary. In contrast to
the 1D boundary, the zero-energy boundary states emerge at the corners of a
nanodisk respecting the mirror symmetries, which are topological as in the
case of the SSH model\cite{SSH}, We cut a crystal along two lines to create
two edges and one corner. These edges are skew-zigzag type to respect the
two mirror symmetries. Actually we analyze the band structure of a nanodisk
respecting the mirror symmetries, which is a 0D crystal. The simplest one is
the diamond structure with four skew-zigzag edges as in Fig.\ref{FigDiamond}(d). 
Corner zero-energy states emerge at the top and bottom lattice sites
for $\left\vert t_{a}\right\vert <t_{b}/2$, as in Fig.\ref{FigDiamond}(e).
When we put one electron into the zero-energy states, the $1/2$ fractional
charge appears at each of the two corners of the diamond. They are absorbed
in metallic phase for $\left\vert t_{a}\right\vert >t_{b}/2 $. These corner
zero-energy states are topological as in the case of the SSH model\cite{SSH}, 
breathing Kagome and breathing Pyrochlore lattices\cite{EzawaPhos}.
Namely, we find that SSH chains are realized at the four edges of the
diamond structure. Consequently, the system is a second-order TI.

(iv) Finally, we study the bulk topological index characterizing the HOTI.
We first investigate the extreme case of $t_{a}=0$. The eigen function for
the valence band is given by $\psi =\left( -e^{ik_{y}},1\right) /\sqrt{2}$.
The Berry connections are obtained as $A_{x}=0$ and $A_{y}=1/2$, which
yields the polarization $p_{x}=0$ and $p_{y}=1/2$. Since the position of the
WC is fixed within one topological phase, we obtain the same result for 
$\left\vert t_{a}\right\vert <t_{b}/2$. The gap closes at $\left\vert
t_{a}\right\vert =t_{b}/2$, and the system becomes semimetallic for 
$\left\vert t_{a}\right\vert >t_{b}/2$ as shown in Fig.\ref{FigDiamond}(d).

\medskip \noindent \textbf{\large Anisotropic diamond lattice}

We proceed to investigate the anisotropic diamond lattice in three
dimensions. We make the five-step arguments.

i) The bulk band spectrum reads
\begin{align}
E^{2}=& t_{a}^{2}+3t_{b}^{2}+2t_{b}^{2}\cos 2\left( k_{x}-k_{y}\right)
+2t_{b}^{2}\cos 2\left( k_{y}-k_{z}\right)   \notag \\
& +2t_{b}^{2}\cos 2\left( k_{z}-k_{x}\right) +2t_{a}t_{b}\cos 2\left(
k_{x}+k_{y}\right)   \notag \\
& +2t_{a}t_{b}\cos 2\left( k_{y}+k_{z}\right) +2t_{a}t_{b}\cos 2\left(
k_{z}+k_{x}\right) .
\end{align}
We show the Fermi surface in Fig.\ref{FigAniBand}(a1)-(d1). The Fermi
surface becomes a loop node for $t_{b}/3<\left\vert t_{a}\right\vert <t_{b}$, 
whose radius shrinks as the ration $\left\vert t_{a}/t_{b}\right\vert $
decreases as in Fig.\ref{FigAniBand}(b1) and (c1). The system becomes an
insulator for $\left\vert t_{a}\right\vert <t_{b}/3$ as in Fig.\ref{FigAniBand}(d1).

ii) We calculate the surface band structure of a thin film in the [111]
direction corresponding to the ($j-1$)D geometry with $j=3$. Zero-energy
partial flat bands appear whose boundary is the projection of the loop node
onto the [111] direction for $t_{b}/3<\left\vert t_{a}\right\vert <t_{b}$:
See Fig.\ref{FigAniBand}(b2) and (c2). It becomes a perfect flat band for 
$\left\vert t_{a}\right\vert <t_{b}/3$ corresponding to the fact that the the
system becomes an insulator: See Fig.\ref{FigAniBand}(d2).

iii) We next calculate the band structure of a diamond prism corresponding
to the ($j-2$)D geometry with $j=3$. As in the case of the thin film, we
find zero-energy partial flat bands for 
$t_{b}/3<\left\vert t_{a}\right\vert <t_{b}$ [Fig.\ref{FigAniBand}(b3), (c3)], 
and zero-energy perfect flat bands for 
$\left\vert t_{a}\right\vert <t_{b}/3$ [Fig.\ref{FigAniBand}(d3)].

iv) We investigate the energy spectrum of the rhombohedron made of the
diamond lattice, which corresponds to the ($j-3$)D geometry with $j=3$: See
Fig.\ref{FigAniDia}. The energy spectrum as a function of $t_{a}/t_{b}$ is
shown in Fig.\ref{FigAniDia}(d). The zero-energy states emerge for 
$|t_{a}|<t_{b}/3$. The emergence of the zero-energy states are naturally
understood by considering the extreme case with $t_{a}=0$. In this case the
two atoms at the top and bottom of the rhombohedron are perfectly isolated
as shown in Fig.\ref{FigAniDia}(a). The $1/2$ fractional charge appears at
each of the two corners of the rhombohedron as shown in Fig.\ref{FigAniDia}(e). 
These zero-energy states are protected by the chiral symmetry and
remain as they are as long as the bulk band gap does not close.

v) We finally analyze the WC. It is easy to derive this at $t_{a}=0$, where
the eigen function for the valence band is given by $\psi =\left(
-e^{i\left( k_{x}+k_{y}+k_{z}\right) },1\right) /\sqrt{2}$. The WC is
calculated as $(1/2,1/2,1/2)$ for $|t_{a}|<t_{b}/3$. The gap closes at 
$\left\vert t_{a}\right\vert =t_{b}/3$, and the system becomes semimetallic
for $\left\vert t_{a}\right\vert >t_{b}/3$, as shown in \ref{FigAniDia}(d).

\medskip \noindent \textbf{\large Discussions}

We have proposed minimal HOTI models. We have argued that phosphorene is a
second-order topological insulator. The construction of a HOTI based on a
nontrivial WC will be applicable to other lattices with dimerizations.

\medskip\noindent\textbf{Acknowledgement}

The author is very much grateful to N. Nagaosa for helpful discussions on
the subject. This work is supported by the Grants-in-Aid for Scientific
Research from MEXT KAKENHI (Grant Nos.JP17K05490 and JP15H05854). This work
is also supported by CREST, JST (JPMJCR16F1).


\begin{thebibliography}{99}
\bibitem{Hasan} M. Z. Hasan and C. L. Kane, Rev. Mod. Phys. \textbf{82},
3045 (2010).

\bibitem{Qi} X.-L. Qi and S.-C. Zhang, Rev. Mod. Phys. \textbf{83}, 1057
(2011).

\bibitem{Fan} F. Zhang, C.L. Kane and E.J. Mele, Phys. Rev. Lett. \textbf{110}, 046404 (2013).

\bibitem{Science} W. A. Benalcazar, B. A. Bernevig, and T. L. Hughes,
10.1126/science.aah6442.

\bibitem{APS} F. Schindler, A. Cook, M. G. Vergniory, and T. Neupert, in APS
March Meeting (2017).

\bibitem{Peng} Y. Peng, Y. Bao, and F. von Oppen, Phys. Rev. B \textbf{95},
235143 (2017).

\bibitem{Lang} J. Langbehn, Y. Peng, L. Trifunovic, F. von Oppen, and P. W.
Brouwer, Phys. Rev. Lett. \textbf{119}, 246401 (2017).

\bibitem{Song} Z. Song, Z. Fang, and C. Fang, Phys. Rev. Lett. \textbf{119},
246402 (2017). (2017).

\bibitem{Bena} W. A. Benalcazar, B. A. Bernevig, and T. L. Hughes, Phys.
Rev. B \textbf{96}, 245115 (2017).

\bibitem{Schin} F. Schindler, A. M. Cook, M. G. Vergniory, Z. Wang, S. S. P.
Parkin, B. A. Bernevig, and T. Neupert, cond-mat/arXiv:1708.03636 (2017).

\bibitem{Liu} M. Lin and T. L. Hughes, arXiv:1708.08457.

\bibitem{EzawaKagome} M. Ezawa, cond-mat/arXiv:1709.08425, Phys. Rev. Lett.
(to be published).

\bibitem{Po} H. C. Po, H. Watanabe and A. Vishwanath, arXiv:1709.06551.

\bibitem{Po2} H. C. Po, H. Watanabe, M. P. Zaletel, and A. Vishwanath, Sci.
Adv. 2 (2016).

\bibitem{Po3} H. C. Po, A. Vishwanath, and H. Watanabe, Nature
Communications 8, 50 (2017).

\bibitem{Brad} B. Bradlyn, L. Elcoro, J. Cano, M. G. Vergniory, Z. Wang, C.
Felser, M. I. Aroyo, and B. A. Bernevig, Nature 547, 298 (2017).

\bibitem{Nanoribbon} M. Ezawa, Phys. Rev. B, \textbf{73}, 045432 (2006).

\bibitem{Nanodisk} M. Ezawa, Phys. Rev. B \textbf{76}, 245415 (2007).

\bibitem{SSH} W. P. Su, J. R. Schrieffer, and A. J. Heeger, Phys. Rev. Lett. 
\textbf{42}, 1698 (1979).

\bibitem{Wunsch} B. Wunsch, F. Guinea and F. Sols, New J. of Phys., 10,
103027 (2008).

\bibitem{Goerbig} G. Montambaux, F. Piechon, J.-N. Fuchs, and M. O. Goerbig,
Phys. Rev. B 80, 153412 (2009).

\bibitem{Pereira} V.M. Pereira,A.H. Castro Neto and N.M.R. Peres, Phys. Rev.
B \textbf{80}, 045401 (2009).

\bibitem{Monta} G. Montambaux F. Pi\'{e}chon, J.-N. Fuchs, and M. O.
Goerbig, Phys. Rev. B \textbf{80}, 153412 (2009).

\bibitem{EzawaPhos} M. Ezawa, New J. Phys. \textbf{16}, 115004 (2014).

\bibitem{Takahashi} R. Takahashi and S. Murakami, Phys. Rev. B 88, 235303
(2013).

\bibitem{Kats} N. Rudenko and M.I. Katsnelson, Phys. Rev. B \textbf{89},
201408 (2014).

\bibitem{Gru} M. M. Grujic, M. Ezawa, M. Z. Tadic, F. M. Peeters, Phys. Rev.
B \textbf{93}, 245413 (2016).
\end{thebibliography}
\end{document}